\documentclass[conference,a4paper]{APSIPA2021}
\usepackage{multirow}
\usepackage[dvips]{graphicx}
\usepackage{amsmath}
\usepackage[psamsfonts]{amssymb}
\usepackage{amsxtra}
\usepackage{threeparttable}

\usepackage[dvips]{hyperref}
\usepackage{breakurl}

\begin{document}

\title{Empirical Study Incorporating\\Linguistic Knowledge on Filled Pauses\\for Personalized Spontaneous Speech Synthesis}

\author{%
\authorblockN{%
Yuta Matsunaga\authorrefmark{1},
Takaaki Saeki\authorrefmark{1},
Shinnosuke Takamichi\authorrefmark{1} and
Hiroshi Saruwatari\authorrefmark{1}
}
\authorblockA{%
\authorrefmark{1}
Graduate School of Information Science and Technology, The University of Tokyo, Japan. 
}
}

\setlength{\tabcolsep}{1mm} 

\maketitle
\thispagestyle{empty}

\begin{abstract}
  We present a comprehensive empirical study for personalized spontaneous speech synthesis on the basis of linguistic knowledge. With the advent of voice cloning for reading-style speech synthesis, a new voice cloning paradigm for human-like and spontaneous speech synthesis is required. We, therefore, focus on personalized spontaneous speech synthesis that can clone both the individual's voice timbre and speech disfluency. Specifically, we deal with filled pauses, a major source of speech disfluency, which is known to play an important role in speech generation and communication in psychology and linguistics. To comparatively evaluate personalized filled pause insertion and non-personalized filled pause prediction methods, we developed a speech synthesis method with a non-personalized external filled pause predictor trained with a multi-speaker corpus. The results clarify the position-word entanglement of filled pauses, i.e., the necessity of precisely predicting positions for naturalness and the necessity of precisely predicting words for individuality on the evaluation of synthesized speech.
\end{abstract}

\section{Introduction} \label{sec:intro}

\begin{figure}[t]
        \centering
        \includegraphics[scale=0.26]{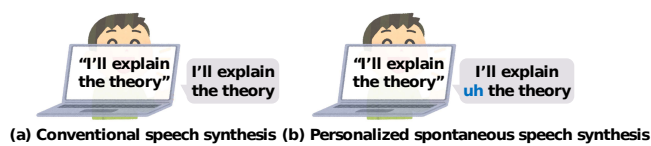}
        \vspace{-4mm}
        \caption{Personalized speech synthesis. Unlike conventional method (left) that reproduces only individual's voice timbre, our method (right) further aims to reproduce individual's speech disfluency.}
        \vspace{-3mm}
        \label{fig:concept}
    \end{figure}

    Speech synthesis aims to artificially synthesize human-like speech. With the rapid development of sequence-to-sequence (seq2seq) models, recent reading-style text-to-speech synthesis can achieve near-human quality~\cite{shen17tacotron2,ren2021fastspeech,weiss21wave,donahue2021endtoend}. Such speech synthesis can highly reproduce speaker individuality, thus has been used to achieve digital voice cloning~\cite{xie21voicecloningchallenge,blaaw19singingvlocecloning,arik18voicecloningfewsample}. With the development of such reading-style speech synthesis, spontaneous speech synthesis has also been studied as a more challenging topic. Compared with reading-style speech, spontaneous speech has a unique characteristic, \textit{speech disfluency}~\cite{schriberg94preliminariesTA}, including repetition, rephrasing, and filled pauses. Spontaneous speech synthesis expresses human-like disfluency and enables us to use speech synthesis for more than just text reading. If spontaneous speech synthesis can be personalized, it should be possible to reproduce individuality beyond conventional voice cloning. Therefore, as shown in Fig.~\ref{fig:concept}, our goal was to achieve \textit{personalized spontaneous speech synthesis} for reproducing not only an individual's voice timbre but also speech disfluency.
    
    Filled pauses (FPs), a type of disfluency, are not relevant to linguistic content but play an important role in spontaneous speech. FPs are inserted into utterances by speakers, and their various roles have been extensively studied in linguistics. Speakers can express planning problems~\cite{levelt83monitoring} and carry out smooth speech communication~\cite{schriberg94preliminariesTA,maclayosgood59hesitation,gravano11turntakingcue} by using FPs. FPs also affect the listener: FPs by a speaker tend to reduce the listening effort required of the listener~\cite{arnold04prednew}. Apart from these studies on the functions of FPs, there have been studies on the individuality of FPs by each speaker. For example, word choice~\cite{schriberg94preliminariesTA,watanabe19japanesefiller} and position~\cite{shriberg96disfluencies} have been reported to change speaker by speaker. To personalize spontaneous speech synthesis, we need to consider all these factors.

    This paper presents a spontaneous speech synthesis method and empirical studies incorporating linguistic knowledge for personalized spontaneous speech synthesis that deals with FPs, with the aim of comparatively evaluating personalized FP insertion (i.e., using a ground-truth FP) and non-personalized FP prediction. To this end, we developed a seq2seq spontaneous speech synthesis model with an external FP predictor. We insert FPs predicted with a prediction model handling a variety of speakers to FP-excluded text and synthesize a speech waveform from the FP-inserted text. We investigated the effect of FP prediction on speech evaluation on the basis of linguistic knowledge of FPs. We evaluated synthetic speech using these functions in terms of not only naturalness (as in AdaSpeech~$3$~\cite{yan2021adaspeech3}) but also speaker individuality and listening effort. Our corpus, audio samples, and open-source implementation are available from our project page\footnote{\url{https://sites.google.com/view/ymatsunaga/publications/fp_synth_22}}. The key contributions of this work are as follows:

    \begin{itemize} \leftskip -2mm \itemsep 1.0mm
        \item We propose a spontaneous speech synthesis method with an external FP predictor and compared a personalized FP insertion with a non-personalized FP prediction.
        \item We constructed an FP word vocabulary and open-source Japanese spontaneous speech corpus (JLecSponSpeech).
        \item Our experimental results clarify the position-word entanglement of FPs, i.e., the necessity of precisely predicting positions for naturalness and the necessity of precisely predicting words for individuality on the evaluation of synthesized speech.
    \end{itemize}

\section{Related work} \label{sec:relatedwork}

    \subsection{Linguistic knowledge on filled pauses.}
        FPs play several roles in spontaneous speech, e.g., speech planning~\cite{schriberg94preliminariesTA,levelt83monitoring} and turn handing~\cite{gravano11turntakingcue}. We summarize the following key linguistic characteristics of FPs that are considered for personalized spontaneous speech synthesis.

        \begin{itemize} \leftskip -2mm \itemsep 1mm
            \item \textbf{Vocabulary:} The vocabulary of FPs is extensive. The two best-known words are ``uh'' and ``um'' (used in AdaSpeech~$3$~\cite{yan2021adaspeech3}), but there are many other lexical and non-lexical words~\cite{brown17listening}. This is also true for other languages~\cite{strassel05chinesefilledpause,zhao05mandarinfilledpause}; Japanese, the target language in this paper, has $160$ different FPs~\cite{hirose06japanesefiller}. 
            \item \textbf{Replaceability:} The roles of each FP word are not strictly separated, and replacing one FP word with another may not change the effect~\cite{yamashita07fillerinfig}.
            \item \textbf{Individuality:} The characteristics of FPs are changed by not only linguistic content~\cite{clark98repeat} but also speaker individuality. Positions~\cite{shriberg96disfluencies} and word choice~\cite{schriberg94preliminariesTA,watanabe19japanesefiller} of FPs differ among speakers.
        \end{itemize}

    \subsection{Prediction and synthesis of filled pauses.}
        Several studies have addressed spontaneous speech synthesis including FPs~\cite{adell08inserteditingterms,dall17spssspontaneous,yan2021adaspeech3}. Adell et al. proposed a disfluent speech synthesis model where editing terms such as FPs are inserted into fluent speech and performed local prosodic modifications~\cite{adell08inserteditingterms}. Dall comprehensibly addressed spontaneous speech synthesis including prediction, synthesis, and evaluation of FPs~\cite{dall17spssspontaneous}. However, these studies did not consider individuality. Yan et al. proposed a multi-speaker speech synthesis model employing spontaneous-style adaptation~\cite{yan2021adaspeech3}; however, they used limited FP vocabulary and did not closely investigate the effect of FP insertion on individuality.    
        Although other studies focused on predicting the positions and words of FPs~\cite{ohta07languagemodelusingfillerprediction,tomalin15latticebasedfillerprediction,yamazaki20blstmfillerprediction}, their effect on synthetic speech remains unknown. In contrast to these studies, we 1) developed a spontaneous speech synthesis method with a newly developed FP vocabulary handling personalization and 2) conducted empirical evaluations by comparing personalized FP insertion and non-personalized FP prediction method in a synthetic speech on the basis of the linguistic knowledge summarized above.

    \subsection{Evaluation of filled pauses.}
        A few studies investigated the effect of the presence of FPs on the personality~\cite{gustafson21personalityinthemix} or individuality~\cite{szekely19conversationalsynthfounddata} of synthetic speech. They compared synthetic speech with and without ground-truth FPs, i.e., individual-specific FPs.
        A study also compared synthetic speech with predicted FPs, ground-truth FPs, and opposite types of ground-truth FPs in terms of plausibility~\cite{szekely19howtotrainfiller}; however, they were not evaluated in terms of individuality, and the FP vocabulary and compared methods of FP insertion were limited.
        We provide deeper insights into the personalization of FP-included spontaneous speech by using position-word decomposition and comparing personalized insertion with non-personalized prediction.

\section{Method} \label{sec:method}

    \subsection{Spontaneous speech synthesis method} \label{ssec:method_model}
    
\begin{figure}[t]
            \centering
            \includegraphics[scale=0.37]{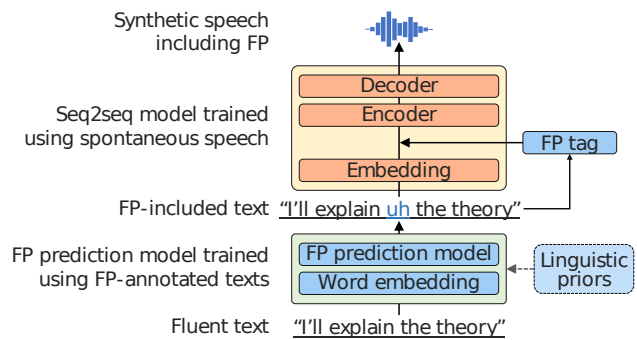}
            \vspace{-3mm}
            \caption{Proposed spontaneous speech synthesis method consisting of two steps: FP prediction and speech synthesis. Linguistic knowledge can be incorporated into FP prediction.}
            \label{fig:method}
            \vspace{-4mm}
        \end{figure}

        We developed two methods: 1) a speaker-dependent speech synthesis model with non-personalized FP prediction and 2) one with personalized FP insertion. We show the former method in Figure~\ref{fig:method}. 
        The FP prediction model consists of two sub-models~\cite{yamazaki20blstmfillerprediction}: bi-directional encoder representations from transformers (BERT)~\cite{devlin19bert} that embeds an input text without FPs (word sequence) and bi-directional long short-time memory (BLSTM)~\cite{graves05blstm} that predicts FPs among input words. 
        Items to be predicted are ``None'' (i.e., FP is not inserted) and the FP words mentioned in Section~\ref{ssec:method_vocab}. 
        We used the non-personalized FP prediction model proposed in a previous study~\cite{matsunaga22fppredgroup}.
        On the other hand, to construct the latter method, we used ground-truth FPs as personalized FPs, instead of predicted FPs.
        The seq2seq speech synthesis model synthesizes spontaneous speech with FPs from an FP-inserted text with the above prediction and insertion methods. In addition to phoneme inputs, we use binary FP tags to indicate whether the current phoneme is an FP. 

        An alternative to the above FP prediction model would be to train a personalized model (i.e., specializing in predicting FPs of a particular speaker); however, adapting an FP prediction model to a particular speaker has been difficult~\cite{matsunaga22fppredgroup}.
        Therefore, we compared personalized FP insertion with ground-truth FPs and non-personalized FP prediction. Our future work will address the personalization of the FP prediction model.
        
        We can incorporate linguistic knowledge of FPs into spontaneous speech synthesis methods. For example, we can consider replaceability (in Section~\ref{sec:relatedwork}), which suggests we can predict FP words less accurately than the positions. In such a case, the prediction model predicts only FP words, given ground-truth FP positions. The details are described in Section~\ref{sec:evaluation}.

    \subsection{Design of filled pause vocabulary} \label{ssec:method_vocab}
        For personalized spontaneous speech synthesis, we need to build an FP word vocabulary to handle spontaneous speech by various speakers. For this purpose, using an FP-annotated text corpus transcribed from multi-speaker spontaneous speech, we exclude any FP words used less frequently ($<$ $20$\%) among all speakers. This creates a vocabulary that consists of $13$ FP words, e.g., ``ee'' and ``ma'', and covers approximately $83$\% on average and at least $53$\% of FPs used by each speaker.

    \subsection{Corpus and annotation} \label{ssec:method_corpus}
        We need to construct a mid-size ($3.5$--$5.0$ hours) spontaneous speech corpus to train the speaker-dependent speech synthesis model. We manually searched for lecture videos on the web and transcribed and annotated texts according to the rules of the Corpus of Spontaneous Japanese (CSJ~\cite{maekawa04csj}). The corpus includes transcribed fluent text as well as FP words, FP tags, and phrase timings. It is available from our project page.

\section{Experimental evaluation} \label{sec:evaluation}

    \subsection{Experimental settings} \label{ssec:setting}
    We used the texts of $137$~speakers from the CSJ~\cite{maekawa04csj} to build an FP word vocabulary and train the FP prediction model. The process is described in Section~\ref{ssec:method_vocab}.
    
    We selected two Japanese speakers whose lecture videos are available on YouTube (see our project page). The audio data were down-sampled at $22.05$~kHz. Professional annotators carried out the annotation. The duration of training data for each speaker was approximately $3.5$--$5.0$ hours. We split the annotated data into training, validation, and test sets. We first manually selected $20$~speech segments for the test set so that all the segments contain multiple FPs with a well-balanced variety of FP words. Each segment consisted of approximately $2$ or $3$~sentences, and the duration per segment was approximately $15$--$30$ seconds. We then randomly selected $512$~breath groups from the remaining data as the validation set.
    
    The architecture of the FP prediction model is based on previous studies~\cite{yamazaki20blstmfillerprediction,matsunaga22fppredgroup}. As described in Section~\ref{ssec:method_model}, the model consists of a BERT and BLSTM. In the training of the BLSTM, we set the number of hidden layers and hidden size to $1$ and $1024$, respectively. We applied gradient clipping with the maximum of the norm set to $0.5$ and set the batch size to $32$. We used the Adam optimizer~\cite{Kingma15adam} with the learning rate set to $1.0\times10^{-5}$. We then trained the model for $60$~k steps.

    The seq2seq speech synthesis model is based on FastSpeech~$2$ architecture~\cite{ren2021fastspeech}. It was first pre-trained using the JSUT corpus (reading-style speech)~\cite{sonobe2017jsut} then fine-tuned using the annotated spontaneous speech. The hyperparameter settings of the model and pretraining followed a published implementation\footnote{\url{https://github.com/ndkgit339/FastSpeech2-filled_pause_speech_synthesis/tree/master/config/example_train}, commit 50c924a.}. We conducted pre-training and fine-tuning for $600$~k and $500$~k steps, respectively. We used accent features in addition to phonemes~\cite{kurihara21prosodicfeats}. We divided spontaneous speech into breath groups for fine-tuning. The phoneme-level binary FP tags were concatenated with phoneme embeddings and then fed to the encoder. The speech was independently synthesized for each breath group and then simply concatenated to make speech segments.

    \subsection{Subjective evaluations} \label{ssec:exp_compara}

    We investigated the effect of FP prediction on synthetic speech. To take into account the replaceability described in Section~\ref{sec:relatedwork}, we separated FP features into FP words and positions.

    \textbf{Evaluation metrics.} We conducted subjective evaluations using synthetic speech. Since our goal was personalized spontaneous speech synthesis, not only naturalness but also speaker individuality was used as evaluation criteria. 
    We also used listening effort, which is an important role of FPs~\cite{arnold04prednew}, as a criterion that means ``which speech sample requires less effort to listen to?''.

    \textbf{Evaluation settings.} Unless otherwise stated, we conducted preference AB and XAB tests. ``X'' is the ground-truth speech (i.e., the speaker's natural speech) in the individuality tests. A total of $30$~listeners participated in each test. Each listener evaluated $10$ and $8$~pairs of speech samples in the preference AB and XAB tests respectively. We randomly selected these samples from the test set.
    
    \begin{table}[t]
\vspace{-0mm}
\footnotesize
\centering
\caption{Preference scores of analysis-synthesized speech\\with and without FPs}
\vspace{-3mm}
\label{tab:eval0}
\begin{tabular}{c|c|cc}
\hline
Criterion                         & Spk. & \begin{tabular}[c]{@{}c@{}}Score\\Analysis-synthesized speech\\without FPs vs. with FPs\end{tabular} & $p$-value                      \\ \hline
\multirow{2}{*}{Naturalness}      & A       & 0.233 vs. \textbf{0.767}                                           & $<$\( 10^{-10} \)  \\
                                  & B       & 0.317 vs. \textbf{0.683}                                           & $<$\( 10^{-10} \) \\ \hline
\multirow{2}{*}{Individuality}    & A       & 0.304 vs. \textbf{0.696}                                         & $<$\( 10^{-10} \)  \\
                                  & B       & 0.313 vs. \textbf{0.687}                                         & $<$\( 10^{-10} \)  \\ \hline
\multirow{2}{*}{Listening effort} & A       & 0.240 vs. \textbf{0.760}                                           & $<$\( 10^{-10} \)  \\
                                  & B       & 0.273 vs. \textbf{0.727}                                           & $<$\( 10^{-10} \)  \\ \hline
\end{tabular}
\vspace{-2mm}
\end{table}

    \textbf{Preliminary evaluation.} 
    First of all, we conducted preliminary evaluations to answer the question, ``Are FPs required for personalized spontaneous speech synthesis?'' Since the evaluation of synthesized speech might be affected by the quality of the synthesis model, we evaluated analysis-synthesized speech with and without ground-truth FPs to exclude the influence of the synthesis model quality.
    We obtained analysis-synthesized speech with FPs by mel-spectrogram analysis and a HiFi-GAN vocoder. We also obtained that without FPs in the same manner; however, we replaced mel-spectrograms of FPs with those of silence with a fixed duration. We set the duration as natural as possible and finally set it to $0.5$~seconds\footnote{There is another way that we replace mel-spectrograms of FPs with those of silence with the same duration as FPs; however, we can obtain more natural speech by replacing with silence with a fixed duration.}. We evaluated those speech samples under the same condition as described above. 
    Table~\ref{tab:eval0} lists the results. The results show that analysis-synthesized speech with ground-truth FPs has significantly higher preference scores in all criteria compared to the case without FPs.
    And, we also evaluated natural speech with and without ground-truth FPs under the same condition. As in the case of analysis-synthesized speech, the results show that natural speech with ground-truth FPs has significantly higher preference scores in all criteria compared to the case without FPs. 
    These results indicate that FPs are required in terms of naturalness, individuality, and listening effort in both analysis-synthesized and natural speech.

    \begin{table}[t]
\vspace{-0mm}
\footnotesize
\centering
\caption{List of methods compared in experiments}
\vspace{-3mm}
\label{tab:method}
\begin{tabular}{cccc}
\hline
Name                      & FP word      & FP position  & Example                               \\ \hline
\textbf{NoW-NoP}     & --           & --           & I explain a theory. \\ \hline
\textbf{RandW-RandP} & Random       & Random       & I explain a theory \textcolor{blue}{uh}. \\ \hline
\textbf{PredW-PredP} & Predicted    & Predicted    & I \textcolor{blue}{uh} explain a theory. \\ \hline
\textbf{PredW-TrueP} & Predicted    & Ground-truth & I explain \textcolor{blue}{um} a theory. \\ \hline
\textbf{TrueW-TrueP} & Ground-truth & Ground-truth & I explain \textcolor{blue}{uh} a theory. \\ \hline
\end{tabular}
\vspace{-2mm}
\end{table}


    \begin{table}[t]
\vspace{-0mm}
\footnotesize
\centering
\caption{Preference scores: NoW-NoP vs. TrueW-TrueP}
\vspace{-3mm}
\label{tab:eval1}
\begin{tabular}{c|c|cc}
\hline
Criterion                         & Spk. & \begin{tabular}[c]{@{}c@{}}Score\\ NoW-NoP vs. TrueW-TrueP\end{tabular} & $p$-value                      \\ \hline
\multirow{2}{*}{Naturalness}      & A       & \textbf{0.660} vs. 0.340                                           & $<$\( 10^{-10} \)  \\
                                  & B       & \textbf{0.563} vs. 0.437                                           & $1.88\times10^{-3}$ \\ \hline
\multirow{2}{*}{Individuality}    & A       & \textbf{0.671} vs. 0.329                                         & $<$\( 10^{-10} \)  \\
                                  & B       & 0.542 vs. 0.458                                         & $6.81\times10^{-2}$  \\ \hline
\multirow{2}{*}{Listening effort} & A       & \textbf{0.660} vs. 0.340                                           & $<$\( 10^{-10} \)  \\
                                  & B       & \textbf{0.560} vs. 0.440                                           & $3.24\times10^{-3}$  \\ \hline
\end{tabular}
\vspace{-2mm}
\end{table}

    \textbf{Evaluated methods.} We compared synthesized speech from FP-inserted text with insertion and prediction methods. Table~\ref{tab:method} lists the methods. 
    The trained seq2seq speech synthesis models were the same among the methods, and only the inputs of those models were different. 
    An input text of \textbf{NoW-NoP} has no FPs, i.e., spontaneous speech without FPs is synthesized. \textbf{RandW-RandP} randomly predicts FPs on the basis of the probability of FP words in the training data of the FP prediction model. \textbf{PredW-PredP} predicts FPs using the trained FP prediction model. \textbf{PredW-TrueP} refers to the FP positions of ground-truth (i.e., annotated data) and predicts only words. \textbf{TrueW-TrueP} refers to both FP positions and words of ground-truth, i.e., the text is completely the same as ground-truth.


    \subsubsection{Filled pause synthesis} \label{sssec:exp1}
    We first investigated how synthesized speech is affected by considering ground-truth FPs. Table~\ref{tab:eval1} lists the results of the evaluation comparing NoW-NoP and TrueW-TrueP.
    We can see that the case without FPs shows better results than that with the ground-truth FPs for naturalness and listening effort and shows a better result for individuality in speaker~A.
    This demonstrates that if the speech synthesis model takes FPs into account, it decreases the naturalness and requires more listening effort compared with the case in which FPs are not taken into account. Moreover, the results suggest that taking FPs into account decreases individuality in certain cases.


    \subsubsection{Filled pause prediction} \label{sssec:exp2}
    To investigate the effectiveness of predicting FPs with PredW-PredP, we compared it and RandW-RandP.
    Table~\ref{tab:eval2} lists the results.
    We can see that predicting FPs significantly improves the quality of synthesized speech in all cases compared with the case with random FPs, indicating that predicting FPs is effective for spontaneous speech synthesis.
    
    \begin{table}[t]
\vspace{-0mm}
\footnotesize
\centering
\caption{Preference scores: PredW-PredP vs. RandW-RandP}
\vspace{-3mm}
\label{tab:eval2}
\begin{tabular}{c|c|cc}
\hline
Criterion                         & Spk. & \begin{tabular}[c]{@{}c@{}}Score\\ PredW-PredP vs. RandW-RandP\end{tabular} & $p$-value                      \\ \hline
\multirow{2}{*}{Naturalness}      & A       & \textbf{0.770} vs. 0.230                                           & $<$\( 10^{-10} \)  \\
                                  & B       & \textbf{0.747} vs. 0.253                                           & $<$\( 10^{-10} \) \\ \hline
\multirow{2}{*}{Individuality}    & A       & \textbf{0.808} vs. 0.192                                           & $<$\( 10^{-10} \)  \\
                                  & B       & \textbf{0.817} vs. 0.183                                           & $<$\( 10^{-10} \)  \\ \hline
\multirow{2}{*}{Listening effort} & A       & \textbf{0.750} vs. 0.250                                           & $<$\( 10^{-10} \)  \\
                                  & B       & \textbf{0.693} vs. 0.307                                           & $<$\( 10^{-10} \)  \\ \hline
\end{tabular}
\vspace{-2mm}
\end{table}
    \begin{table}[t]
\vspace{-0mm}
\footnotesize
\centering
\caption{Preference scores: PredW-PredP vs. TrueW-TrueP}
\vspace{-3mm}
\label{tab:eval3}
\begin{tabular}{c|c|cc}
\hline
Criterion                         & Spk. & \begin{tabular}[c]{@{}c@{}}Score\\ PredW-PredP vs. TrueW-TrueP\end{tabular} & $p$-value                      \\ \hline
\multirow{2}{*}{Naturalness}      & A       & 0.470 vs. 0.530                                           & $1.42\times10^{-1}$  \\
                                  & B       & 0.457 vs. \textbf{0.543}                                           & $3.38\times10^{-2}$ \\ \hline
\multirow{2}{*}{Individuality}    & A       & 0.442 vs. \textbf{0.558}                                           & $1.05\times10^{-2}$  \\
                                  & B       & 0.350 vs. \textbf{0.650}                                           & $<$\( 10^{-10} \)  \\ \hline
\multirow{2}{*}{Listening effort} & A       & 0.487 vs. 0.513                                           & $5.14\times10^{-1}$  \\
                                  & B       & 0.433 vs. \textbf{0.567}                                           & $1.06\times10^{-3}$  \\ \hline
\end{tabular}
\vspace{-2mm}
\end{table}

    \subsubsection{Prediction of filled pause positions and words} \label{sssec:exp3}
    We compared the synthetic speech quality when the FP position and word are predicted and when the ground-truth FPs are used.
    Table~\ref{tab:eval3} lists the results for PredW-PredP vs. TrueW-TrueP.
    TrueW-TrueP scored higher on individuality for both speakers, indicating that predicting FPs more precisely improves the individuality of synthesized speech with FPs.
    TrueW-TrueP scored significantly higher on naturalness and listening effort for speaker~B. The low score of naturalness might be because the tendencies of predicted FPs are different from those of the training data of the speech synthesis model. The difference in the scores of listening effort between speakers might be affected by the naturalness of synthesized speech and the speaking rates of the speakers. The latter has been reported to affect listening comprehension~\cite{griffiths90speechrate}. Speaker~B speaks faster than speaker~A ($227$ and $244$ words per minute, respectively), and we expect that precise prediction of FPs is required in faster speech.
    
    
    \subsubsection{Prediction of filled pause words} \label{sssec:exp4}
    To investigate the importance of predicting FP words more precisely, we compared PredW-TrueP and TrueW-TrueP. Table~\ref{tab:eval4} lists the results.
    We can see that the two methods had no significant difference in naturalness and listening effort, indicating that predicting FP ``words'' more precisely cannot significantly improve the naturalness and listening effort of synthetic speech with FPs. The replaceability of the effect of FPs might be also true on synthesized speech because replacing ground-truth words of FPs with predicted words did not significantly lower the scores of listening effort.
    On the other hand, TrueW-TrueP significantly scored higher in individuality for speaker~A, indicating that predicting FP words more precisely improves the individuality of synthesized speech for a certain speaker.
    Such a difference in scores between speakers might be because the accuracy of predicting FP words was lower for speaker~A than speaker~B~\cite{matsunaga22fppredgroup} and more accurate FP words with ground-truth improve individuality.

    \begin{table}[t]
\vspace{-0mm}
\footnotesize
\centering
\caption{Preference scores: PredW-TrueP vs. TrueW-TrueP}
\vspace{-3mm}
\label{tab:eval4}
\begin{tabular}{c|c|cc}
\hline
Criterion                         & Spk. & \begin{tabular}[c]{@{}c@{}}Score\\ PredW-TrueP vs. TrueW-TrueP\end{tabular} & $p$-value                      \\ \hline
\multirow{2}{*}{Naturalness}      & A       & 0.470 vs. 0.530                                           & $1.42\times10^{-1}$  \\
                                  & B       & 0.493 vs. 0.507                                           & $7.44\times10^{-1}$ \\ \hline
\multirow{2}{*}{Individuality}    & A       & 0.454 vs. \textbf{0.546}                                           & $4.47\times10^{-2}$  \\
                                  & B       & 0.496 vs. 0.504                                           & $8.56\times10^{-1}$  \\ \hline
\multirow{2}{*}{Listening effort} & A       & 0.463 vs. 0.537                                           & $7.27\times10^{-2}$ \\
                                  & B       & 0.527 vs. 0.473                                           & $1.92\times10^{-1}$  \\ \hline
\end{tabular}
\vspace{-2mm}
\end{table}
    
    \subsubsection{Prediction of filled pause positions} \label{sssec:exp5}
    To investigate the effect of the FP position on synthetic speech, we compared PredW-PredP and PredW-TrueP.
    Table~\ref{tab:eval5} lists the results. We can see that PredW-TrueP significantly scored higher in naturalness for both speakers, indicating that predicting FP ``positions'' more precisely improve the naturalness of synthesized speech. On the other hand, there was no significant difference between the two methods in individuality and listening effort, indicating that predicting only FP positions more precisely cannot significantly improve the individuality and listening effort of synthetic speech with FPs. 
    
    \begin{table}[t]
\vspace{-0mm}
\footnotesize
\centering
\caption{Preference scores: PredW-PredP vs. PredW-TrueP}
\vspace{-3mm}
\label{tab:eval5}
\begin{tabular}{c|c|cc}
\hline
Criterion                         & Spk. & \begin{tabular}[c]{@{}c@{}}Score\\ PredW-PredP vs. PredW-TrueP\end{tabular} & $p$-value                      \\ \hline
\multirow{2}{*}{Naturalness}      & A       & 0.437 vs. \textbf{0.563}                                           & $1.88\times10^{-3}$  \\
                                  & B       & 0.423 vs. \textbf{0.577}                                           & $1.63\times10^{-4}$ \\ \hline
\multirow{2}{*}{Individuality}    & A       & 0.542 vs. 0.458                                           & $6.81\times10^{-2}$  \\
                                  & B       & 0.479 vs. 0.521                                           & $3.62\times10^{-1}$  \\ \hline
\multirow{2}{*}{Listening effort} & A       & 0.470 vs. 0.530                                           & $1.42\times10^{-1}$  \\
                                  & B       & 0.503 vs. 0.497                                           & $8.71\times10^{-1}$  \\ \hline
\end{tabular}
\vspace{-2mm}
\end{table}

    \subsubsection{Absolute scales}
    Finally, we evaluated the methods in terms of absolute scales. We removed PredW-TrueP, which had only slight differences from TrueW-TrueP as shown in Table~\ref{tab:eval4}, and added ground-truth speech (i.e., natural speech). We conducted five-scale mean opinion score (MOS) tests for naturalness and listening effort and a five-scale degradation MOS (DMOS) test using ground-truth speech as a reference for individuality. We synthesized spontaneous speech from a short-form sentence, not from a segment as in the above evaluations. This is for providing insights into spontaneous speech synthesis, compared with reading-style speech synthesis that is often evaluated using a short-form sentence. A total of $100$~listeners participated in each test. Each listener evaluated $15$ and $12$~speech samples for the MOS and DMOS tests respectively. Table~\ref{tab:mos} lists the results.
    The naturalness scores of synthesized spontaneous speech without FPs (NoW-NoP) were around $3.3$--$3.6$. Considering those of synthesized reading-style speech based on FastSpeech~$2$ trained using LJSpeech (a corpus including slight reverberations, the same as ours) are approximately $3.5$~\cite{hayashi21espnet2}, the quality of our spontaneous speech synthesis is comparable or slightly inferior to recent reading-style speech synthesis. We can see that other results are consistent with those of previous preference tests.
    
    \begin{table}[t]
\vspace{-0mm}
\centering
\caption{Mean opinion score of synthetic speech}
\vspace{-3mm}
\label{tab:mos}
\footnotesize
\begin{tabular}{c|c|cc}
\hline
\multicolumn{1}{c|}{\multirow{2}{*}{Criterion}}      & \multirow{2}{*}{Method} & \multicolumn{2}{c}{Mean $\pm$ 95\% conf interval} \\ \cline{3-4} 
\multicolumn{1}{l|}{}             &                  & \multicolumn{1}{c|}{Spk.~A} & Spk.~B \\ \hline
\multirow{5}{*}{Naturalness}      & NoW-NoP     & \multicolumn{1}{l|}{3.313 $\pm$ 0.139} & 3.619 $\pm$ 0.151 \\
                                  & TrueW-TrueP & \multicolumn{1}{l|}{3.018 $\pm$ 0.140} & 3.366 $\pm$ 0.152 \\
                                  & PredW-PredP & \multicolumn{1}{l|}{2.880 $\pm$ 0.137} & 3.179 $\pm$ 0.159 \\
                                  & RandW-RandP & \multicolumn{1}{l|}{2.157 $\pm$ 0.151} & 2.545 $\pm$ 0.182 \\
                                  & Ground-truth      & \multicolumn{1}{l|}{4.133 $\pm$ 0.149} & 4.313 $\pm$ 0.141 \\ \hline
\multirow{4}{*}{Individuality}    & NoW-NoP     & \multicolumn{1}{l|}{3.167 $\pm$ 0.159} & 3.286 $\pm$ 0.186 \\
                                  & TrueW-TrueP & \multicolumn{1}{l|}{3.057 $\pm$ 0.156} & 3.373 $\pm$ 0.181 \\
                                  & PredW-PredP & \multicolumn{1}{l|}{2.937 $\pm$ 0.151} & 3.008 $\pm$ 0.187 \\
                                  & RandW-RandP & \multicolumn{1}{l|}{2.167 $\pm$ 0.154} & 2.460 $\pm$ 0.186 \\ \hline
\multirow{5}{*}{Listening effort} & NoW-NoP     & \multicolumn{1}{l|}{3.213 $\pm$ 0.152} & 3.405 $\pm$ 0.177 \\
                                  & TrueW-TrueP & \multicolumn{1}{l|}{3.115 $\pm$ 0.152} & 3.373 $\pm$ 0.168 \\
                                  & PredW-PredP & \multicolumn{1}{l|}{2.868 $\pm$ 0.158} & 3.135 $\pm$ 0.164 \\
                                  & RandW-RandP & \multicolumn{1}{l|}{2.368 $\pm$ 0.163} & 2.635 $\pm$ 0.200 \\
                                  & Ground-truth      & \multicolumn{1}{l|}{3.822 $\pm$ 0.165} & 3.937 $\pm$ 0.162 \\ \hline
\end{tabular}
\vspace{-4mm}
\end{table}

    \subsection{Analysis}
    This section provides a summary discussion of the evaluations presented in Section~\ref{ssec:exp_compara}.
    The evaluations of analysis-synthesized speech indicate that FP insertion improved evaluated scores on all criteria. 
    However, as shown in Table~\ref{tab:eval1}, inserting synthesized FPs lowered the scores, corresponding to the results of a previous study~\cite{gustafson21personalityinthemix}. This indicates that the quality of speech synthesis with FPs is still low.
    Moreover, the results of the preference scores of naturalness in Table~\ref{tab:eval4} and~\ref{tab:eval5} suggest that accurately predicting FP positions, not FP words, is required for improving the naturalness of synthesized spontaneous speech with FPs. On the other hand, those of individuality in the tables suggest that accurately predicting FP words can improve individuality for a speaker with less accurate prediction, while accurately predicting only FP positions cannot. Thus, accurate prediction of FP words, not only positions, is required for personalizing speech synthesis with FPs. 
    Therefore, we will work on improving spontaneous speech synthesis with and without FPs and the prediction of FP words and positions.

\section{Conclusion} \label{sec:conclusion}
    We proposed a spontaneous speech synthesis method including filled pause prediction. With linguistic knowledge as a basis, we explored the effect of filled pause prediction and identified several potential directions for the personalization of spontaneous speech synthesis. Our future work will focus on improving the basic performance of spontaneous speech synthesis including filled pauses and the automatic development of spontaneous speech corpora.

\section*{Acknowledgment}
This work was supported by JST, Moonshot R\&D Grant Number JPMJPS2011.

\bibliographystyle{IEEEtran}
\bibliography{mybib}

\begin{thebibliography}{10}
\providecommand{\url}[1]{#1}
\csname url@samestyle\endcsname
\providecommand{\newblock}{\relax}
\providecommand{\bibinfo}[2]{#2}
\providecommand{\BIBentrySTDinterwordspacing}{\spaceskip=0pt\relax}
\providecommand{\BIBentryALTinterwordstretchfactor}{4}
\providecommand{\BIBentryALTinterwordspacing}{\spaceskip=\fontdimen2\font plus
\BIBentryALTinterwordstretchfactor\fontdimen3\font minus
  \fontdimen4\font\relax}
\providecommand{\BIBforeignlanguage}[2]{{%
\expandafter\ifx\csname l@#1\endcsname\relax
\typeout{** WARNING: IEEEtran.bst: No hyphenation pattern has been}%
\typeout{** loaded for the language `#1'. Using the pattern for}%
\typeout{** the default language instead.}%
\else
\language=\csname l@#1\endcsname
\fi
#2}}
\providecommand{\BIBdecl}{\relax}
\BIBdecl

\bibitem{shen17tacotron2}
J.~{Shen}, R.~{Pang}, R.~J. {Weiss}, M.~{Schuster}, N.~{Jaitly}, Z.~{Yang},
  Z.~{Chen}, Y.~{Zhang}, Y.~{Wang}, R.~{Skerrv-Ryan}, R.~A. {Saurous},
  Y.~{Agiomvrgiannakis}, and Y.~{Wu}, ``{Natural TTS Synthesis by Conditioning
  WaveNet on MEL Spectrogram Predictions},'' in \emph{Proc. ICASSP}, Apr. 2018,
  pp. 4779--4783.

\bibitem{ren2021fastspeech}
Y.~Ren, C.~Hu, X.~Tan, T.~Qin, S.~Zhao, Z.~Zhao, and T.-Y. Liu, ``{FastSpeech
  2: Fast and High-Quality End-to-End Text to Speech},'' \emph{arXiv preprint
  arXiv:2006.04558}, 2020.

\bibitem{weiss21wave}
R.~J. Weiss, R.~Skerry-Ryan, E.~Battenberg, S.~Mariooryad, and D.~P. Kingma,
  ``{Wave-Tacotron: Spectrogram-Free End-to-End Text-to-Speech Synthesis},'' in
  \emph{Proc. ICASSP}, June 2021, pp. 5679--5683.

\bibitem{donahue2021endtoend}
J.~Donahue, S.~Dieleman, M.~Binkowski, E.~Elsen, and K.~Simonyan, ``{End-to-end
  Adversarial Text-to-Speech},'' in \emph{Proc. ICLR}, May 2021.

\bibitem{xie21voicecloningchallenge}
Q.~Xie, X.~Tian, G.~Liu, K.~Song, L.~Xie, Z.~Wu, H.~Li, S.~Shi, H.~Li, F.~Hong,
  H.~Bu, and X.~Xu, ``{The Multi-Speaker Multi-Style Voice Cloning Challenge
  2021},'' in \emph{Proc. ICASSP}, Jun. 2021, pp. 8613--8617.

\bibitem{blaaw19singingvlocecloning}
M.~Blaauw, J.~Bonada, and R.~Daido, ``{Data Efficient Voice Cloning for Neural
  Singing Synthesis},'' in \emph{Proc. ICASSP}, May 2019, pp. 6840--6844.

\bibitem{arik18voicecloningfewsample}
S.~Arik, J.~Chen, K.~Peng, W.~Ping, and Y.~Zhou, ``{Neural Voice Cloning with a
  Few Samples},'' in \emph{Advances in NeurIPS}, Dec. 2018, pp.
  10\,019--10\,029.

\bibitem{schriberg94preliminariesTA}
S.~Elisabeth, ``{Preliminaries to a Theory of Speech Disfluencies},''
  \emph{Unpublished PhD dissertation, University of California, Berkeley},
  1994.

\bibitem{levelt83monitoring}
W.~J. Levelt, ``{Monitoring and self-repair in speech},'' \emph{Cognition},
  vol.~14, no.~1, pp. 41--104, 1983.

\bibitem{maclayosgood59hesitation}
H.~Maclay and C.~E. Osgood, ``{Hesitation Phenomena in Spontaneous English
  Speech},'' \emph{WORD}, vol.~15, no.~1, pp. 19--44, 1959.

\bibitem{gravano11turntakingcue}
A.~Gravano and J.~Hirschberg, ``{Turn-taking cues in task-oriented dialogue},''
  \emph{Computer Speech \& Language}, vol.~25, no.~3, pp. 601--634, 2011.

\bibitem{arnold04prednew}
J.~E. Arnold, M.~K. Tanenhaus, R.~J. Altmann, and M.~Fagnano, ``{The Old and
  Thee, uh, New: Disfluency and Reference Resolution},'' \emph{Psychological
  Science}, vol.~15, no.~9, pp. 578--582, 2004.

\bibitem{watanabe19japanesefiller}
M.~Watanabe and Y.~Shirahata, ``{Factors Related to Probabilities of
  Clause-Internal ``Ee'', ``Anoo'' and ``Maa'' in Simulated Public Speaking of
  CSJ},'' in \emph{Proc. Language Resources Workshop}, Sep. 2019, pp. 359--367,
  in Japanese.

\bibitem{shriberg96disfluencies}
E.~Shriberg, ``{Disfluencies in switchboard},'' in \emph{Proc. ICSLP}, Oct.
  1996, pp. 11--14.

\bibitem{yan2021adaspeech3}
Y.~Yan, X.~Tan, B.~Li, G.~Zhang, T.~Qin, S.~Zhao, Y.~Shen, W.-Q. Zhang, and
  T.-Y. Liu, ``{Adaptive Text to Speech for Spontaneous Style},'' in
  \emph{Proc. INTERSPEECH}, Aug. 2021, pp. 4668--4672.

\bibitem{brown17listening}
G.~Brown, \emph{{Listening to Spoken English}}, ser. Applied Linguistics and
  Language Study.\hskip 1em plus 0.5em minus 0.4em\relax Taylor \& Francis,
  2017.

\bibitem{strassel05chinesefilledpause}
S.~Strassel, J.~Kolár, Z.~Song, L.~Barclay, and M.~Glenn, ``{Structural
  metadata annotation: moving beyond English},'' in \emph{Proc. INTERSPEECH},
  Sep. 2005, pp. 1545--1548.

\bibitem{zhao05mandarinfilledpause}
Y.~Zhao and D.~Jurafsky, ``{A preliminary study of Mandarin filled pauses},''
  in \emph{Proc. DiSS}, Sep. 2005.

\bibitem{hirose06japanesefiller}
K.~Hirose, Y.~Abe, and N.~Minematsu, ``{Detection of fillers using prosodic
  features in spontaneous speech recognition of Japanese},'' in \emph{Proc.
  Speech Prosody}, May 2006, p. paper 187.

\bibitem{yamashita07fillerinfig}
K.~Yamashita and E.~Mizukami, ``{Using Fillers as Mental Makers: Effects of
  Familiarity, Modality, and Task Difficulty in Describing the Figure},''
  \emph{Journal of Natural Language Processing}, vol.~14, no.~3, pp. 39--60,
  2007, in Japanese.

\bibitem{clark98repeat}
H.~H. Clark and T.~Wasow, ``{Repeating Words in Spontaneous Speech},''
  \emph{Cognitive Psychology}, vol.~37, no.~3, pp. 201--242, 1998.

\bibitem{adell08inserteditingterms}
J.~Adell, A.~Bonafonte, and D.~Escudero-Mancebo, ``{On the generation of
  synthetic disfluent speech: local prosodic modifications caused by the
  insertion of editing terms},'' in \emph{Proc. INTERSPEECH}, Sep. 2008, pp.
  2278--2281.

\bibitem{dall17spssspontaneous}
R.~Dall, ``{Statistical Parametric Speech Synthesis Using Conversational Data
  and Phenomena},'' \emph{PhD dissertation of the University of Edinburgh},
  2017.

\bibitem{ohta07languagemodelusingfillerprediction}
K.~Ohta, M.~Tsuchiya, and S.~Nakagawa, ``{Construction of spoken language model
  including fillers using filler prediction model},'' in \emph{Proc.
  INTERSPEECH}, Aug. 2007, pp. 1489--1492.

\bibitem{tomalin15latticebasedfillerprediction}
M.~Tomalin, M.~Wester, R.~Dall, B.~Byrne, and S.~King, ``{A lattice-based
  approach to automatic filled pause insertion},'' in \emph{Proc. DiSS}, Aug.
  2015.

\bibitem{yamazaki20blstmfillerprediction}
Y.~Yamazaki, Y.~Chiba, T.~Nose, and A.~Ito, ``{Filler Prediction Based on
  Bidirectional LSTM for Generation of Natural Response of Spoken Dialog},'' in
  \emph{Proc. GCCE}, Oct. 2020, pp. 360--361.

\bibitem{gustafson21personalityinthemix}
J.~Gustafson, J.~Beskow, and E.~Szekely, ``{Personality in the mix -
  investigating the contribution of fillers and speaking style to the
  perception of spontaneous speech synthesis},'' in \emph{Proc. 11th ISCA SSW},
  Aug. 2021, pp. 48--53.

\bibitem{szekely19conversationalsynthfounddata}
Éva Székely, G.~E. Henter, J.~Beskow, and J.~Gustafson, ``{Spontaneous
  Conversational Speech Synthesis from Found Data},'' in \emph{Proc.
  INTERSPEECH}, Sep. 2019, pp. 4435--4439.

\bibitem{szekely19howtotrainfiller}
Éva Székely, G.~{Eje Henter}, J.~Beskow, and J.~Gustafson, ``{How to train
  your fillers: uh and um in spontaneous speech synthesis},'' in \emph{Proc.
  10th ISCA SSW}, Sep. 2019, pp. 245--250.

\bibitem{devlin19bert}
J.~Devlin, M.-W. Chang, K.~Lee, and K.~Toutanova, ``{BERT: Pre-training of Deep
  Bidirectional Transformers for Language Understanding},'' \emph{arXiv
  preprint arXiv:1810.04805}, 2019.

\bibitem{graves05blstm}
A.~Graves and J.~Schmidhuber, ``{Framewise phoneme classification with
  bidirectional LSTM and other neural network architectures},'' \emph{Neural
  Networks}, vol.~18, no.~5, pp. 602--610, 2005.

\bibitem{matsunaga22fppredgroup}
Y.~Matsunaga, T.~Saeki, S.~Takamichi, and H.~Saruwatari, ``{Personalized
  Filled-pause Generation with Group-wise Prediction Models},'' in \emph{Proc.
  LREC}, Jun. 2022, pp. 385--392.

\bibitem{maekawa04csj}
K.~Maekawa, ``{Corpus of Spontaneous Japanese : its design and evaluation},''
  in \emph{Proc. SSPR}, Apr. 2003, pp. 7--12.

\bibitem{Kingma15adam}
D.~P. Kingma and J.~Ba, ``{Adam: A Method for Stochastic Optimization},''
  \emph{arXiv preprint arXiv:1412.6980}, 2014.

\bibitem{sonobe2017jsut}
R.~Sonobe, S.~Takamichi, and H.~Saruwatari, ``{JSUT corpus: free large-scale
  Japanese speech corpus for end-to-end speech synthesis},'' \emph{arXiv
  preprint arXiv:1711.00354}, 2017.

\bibitem{kurihara21prosodicfeats}
K.~Kurihara, N.~Seiyama, and T.~Kumano, ``{Prosodic Features Control by Symbols
  as Input of Sequence-to-Sequence Acoustic Modeling for Neural TTS},''
  \emph{IEICE Transactions on Information and Systems}, vol. E104.D, no.~2, pp.
  302--311, 2021.

\bibitem{griffiths90speechrate}
R.~Griffiths, ``{Speech Rate and NNS Comprehension: A Preliminary Study in
  Time-Benefit Analysis},'' \emph{Language Learning}, vol.~40, no.~3, pp.
  311--336, 1990.

\bibitem{hayashi21espnet2}
T.~{Hayashi}, R.~{Yamamoto}, T.~{Yoshimura}, P.~{Wu}, J.~{Shi}, T.~{Saeki},
  Y.~{Ju}, Y.~{Yasuda}, S.~{Takamichi}, and S.~{Watanabe}, ``{ESPnet2-TTS:
  Extending the Edge of TTS Research},'' \emph{arXiv preprint
  arXiv:2110.07840}, 2021.

\end{thebibliography}

\end{document}